\documentclass[preprint]{revtex4-1}
\usepackage{amsmath}
\usepackage{amsfonts}
\usepackage{graphicx}
\usepackage{bbold}
\usepackage{csquotes}
\usepackage[font=scriptsize,labelfont=bf]{caption}

\setcitestyle{authoryear}

\begin{document}

\title{Collapse models and spacetime symmetries}
\date{\today}
\author{Daniel J.~Bedingham}
\email{daniel.bedingham@philosophy.ox.ac.uk}
\affiliation{Faculty of Philosophy, University of Oxford, OX2 6GG, United Kingdom.}

\begin{abstract}
A picture of dynamical collapse of the wave function which is relativistic and time symmetric is presented. The part of the model which exhibits these features is the set of collapse outcomes. These play the role of matter distributed in space and time. It is argued that the dynamically collapsing quantum state, which is both foliation dependent and follows a time-asymmetric dynamics, is not fundamental: it represents a state of information about the past matter distribution for the purpose of estimating the future matter distribution. It is also argued from the point of view of collapse models that both special and general relativistic considerations point towards a discrete spacetime structure and that gravity may not need to be quantised to give a theory that is consistent with quantum matter.
\end{abstract}
\maketitle

\section{Introduction}
\subsection{Destination}
Consider the following description of the physical world:
\begin{displayquote}
The world is described by a definite-valued distribution of matter in space and time. The matter densities assigned to points in the future with respect to some time slice cannot be predicted with certainly. The best that can be done is to estimate these values on the basis of information about matter densities at points in the past, and to use {\it standard methods of Bayesian inference} to update our estimates as new information about the distribution of matter is acquired. We find good agreement with quantum theory. Furthermore, our description is grounded in definite local features---the distribution of matter in space and time. The uncertainty which characterises quantum behaviour results from the uncertain future values of matter density given information about the way that matter was distributed in the past. 
\end{displayquote}

This may not sound like a model of objective collapse of the wave function, yet we aim to show that this is the natural way to understand the meaning of collapse models in light of the symmetries that we expect of their dynamics under transformations of the spacetime coordinates. In order to get there we must tell the story from the beginning.

\subsection{Journey}
In the orthodox picture of quantum theory our only way to access the quantum world is via the results of measurements. It is only via the results of measurements that we can infer the form of the wave function which cannot be observed directly. Despite this lack of direct experience, the ability of the wave function to undergo interference effects which appear in the precisely determined distribution of measurement results leads many to believe that the wave function is a real physical object (see also \citet{pusey2012reality}).

The orthodox quantum rules determine that the wave function should be updated following a measurement in correspondence with the observed outcome. Treating this as a real physical collapse process acting on a real physical wave function raises the question of what should count as a measurement capable of triggering it---this is the infamous measurement problem \citep{albert2009quantum,bell1981quantum,bell1990against}.
The original idea behind the solution to this problem offered by dynamical collapse models \citep{bassi2003dynamical,bassi2013models} was to give the collapse process a firm foundation by having it obey a well-defined law. The essential feature is that it happens, not in response to the occurrence of a measurement, but spontaneously as a result of a mathematically expressed rule. For any sensible dynamical collapse model, it should be such that it reproduces the overall effect of the collapse of the wave function in situations that are unambiguously quantum measurements, that it has little effect on the usual Schr\"odinger dynamics of micro systems, and that, since it is supposed to be a description of matter on all scales, it reduces to a classical description of macro objects.

The fact that it is possible to find such a model is remarkable \citep{ghirardi1986unified,pearle1989combining,ghirardi1990markov}. It offers the chance to solidify the foundations of quantum theory and resolve the measurement problem in a way that adheres to orthodox intuitions. But it may not persuade everyone. Many believe that the collapse process should properly be viewed as a process of inference for which it is natural to regard the wave function as a state of information, and that this should be taken into account in the foundations. Indeed, the point will be stressed in this paper that collapse models can be viewed in this way, where the wave function is demoted from real physical object to a device for encoding information about objective facts. Furthermore, this is a view which is strongly motivated by concerns about the symmetries of the collapse dynamics under transformations of the spacetime coordinates. 

In collapse models, as will be explicitly described, when a collapse takes place, it is formally equivalent in terms of its effect on the wave function to the effect of a quantum measurement. Associated to the collapse is a random variable which plays the role of a measurement outcome. Just as in the orthodox picture where the measurement outcomes give our only direct physical access to the quantum world, it is natural to treat the collapse outcomes as the part of the model which corresponds to real physical events at definite points in space and time. This view with primitive status given to the collapse outcomes rather than the wave function is close to the way in which collapse models were understood by Bell \citep{bell2001there}. The chance of any particular outcome is determined by the wave function, but the wave function itself is determined by the past collapse outcomes. This means that the wave function can be seen as a useful intermediary whose job is to process information about collapses that have occurred in the past in order to estimate future collapse outcomes. Something that we aim to clarify here is that the wave function plays the role of a probability distribution for future collapse outcomes in surprisingly conventional way.

In the terminology of \citet{goldstein1998quantum,goldstein1998quantum2} and 
\citet{allori2012metaphysics,allori2013primitive,allori2015primitive}, the collapse outcomes are to be regarded as the {\it primitive ontology} (PO). The PO is the part of a theory which provides a picture of the distribution of matter in space and time, allowing for a straightforward comparison between the theory and the world of our experience. In the context of collapse models (and in particular the Ghirardi-Rimini-Weber (GRW) model \citep{ghirardi1986unified}---see below) treating the collapse outcomes as the PO has come to be known as the flash ontology.

The generic structure of collapse models will be presented in Section \ref{S2} and it will be demonstrated that a collapse of the wave function is nothing more than a Bayesian updating in response to information about the present distribution of matter. It will also be shown that, under given conditions, this structure has time reversal symmetry at the level of collapse outcomes. This is to be contrasted with the quantum state description which is clearly time asymmetric since it is shaped by collapses in the past but not in the future. In Section \ref{S3} the structure of specially relativistic dynamical collapse is presented where it is shown how the probability of a given set of collapse outcomes in a given spacetime region can be Lorentz invariant and independent of spacetime foliation. A quantum state history by contrast must be defined with reference to a spacetime foliation. If the quantum state were the basis of local physical properties this feature would be troublesome (local features should not depend on an arbitrary choice of time slice). However, with primitive status given to collapse outcomes and the quantum state representing maximal information about past collapses the hypersurface dependence of is the quantum state is natural, since a hypersurface is needed to demarcate the past-future boundary. It is also argued that considerations of energy conservation point towards a discrete spacetime structure. In Section \ref{S4} this view is strengthened by the possibility that it opens a way to consistently describe classical gravity and quantum matter.

\section{Generic non-relativistic structure}
\label{S2}

For all the variation in proposed dynamical collapse models there is a common underlying structure which we present here \citep{bedingham2015time,bedingham2016time}. There is a quantum state or wave function $|\psi_t\rangle$ which, if there is no collapse event between times $s$ and $t>s$, evolves in time satisfying the Schr\"odinger equation with solution
\begin{align}
|\psi_t\rangle = \hat{U}(t-s)|\psi_s\rangle,
\label{eq:CM1}
\end{align}
where $\hat{U}(t) = e^{-i\hat{H}t}$ is the usual unitary time evolution operator for a Hamiltonian $\hat{H}$. At certain times governed by the model there is a collapse event (we assume for now that these times are discrete; they may be randomly distributed or periodic and deterministic). The collapse events involve a sudden change to the wave function. A collapse at time $t$ is described by 
\begin{align}
|\psi_t\rangle \rightarrow |\psi_{t+}\rangle = \hat{J}(z_t)|\psi_t\rangle,
\label{eq:CM2}
\end{align}
where $\hat{J}$ is the collapse operator ($J$ for ``jump") and $z_t$ is a random variable upon which it depends. The chance of realising a particular $z_t$ is given by the probability distribution
\begin{align}
P(z_t|\psi_t\rangle = \frac{\langle \psi_{t+}|\psi_{t+}\rangle}{\langle \psi_{t}|\psi_{t}\rangle}
=\frac{\langle \psi_{t}|\hat{J}^{\dagger}(z_t)\hat{J}(z_t)|\psi_{t}\rangle}{\langle \psi_{t}|\psi_{t}\rangle}.
\label{eq:CM3}
\end{align}
This means that the state immediately prior to the collapse event determines the chance of $z_t$. Finally, in order for the probability distribution to be normalised we require the completeness condition
\begin{align}
\int dz \hat{J}^{\dagger}(z)\hat{J}(z) = \hat{\mathbb{1}}.
\label{eq:CM4}
\end{align}
The last four equations are the essence of the idea of dynamical collapse. A specific model corresponds to a particular choice for $\hat{J}$ and for the sequence of times at which the collapses occur. Continuous models described by stochastic differential equations can be understood as the limit case of this construction in which the collapses happen at each moment (this requires a suitable weakening of the effect of each collapse as the rate of collapses is increased). The choice of collapse operator commits to a definite basis (or set of bases) within which collapse takes place. 

The two most well-known models are the Ghirardi-Rimini-Weber (GRW) model \citep{ghirardi1986unified} and the continuous spontaneous localisation (CSL) model \citep{pearle1989combining,ghirardi1990markov}. For the GRW model we have
\begin{align}
\hat{J}({\bf z}_t) = \left(\frac{\alpha}{\pi}\right)^{3/4} \exp\left\{-\frac{\alpha}{2}(\hat{\bf x}-{\bf z}_t)^2\right\},
\end{align}
where $\hat{\bf x}$ is the position operator for an individual constituent particle and $\alpha$ is a constant. The effect of the collapse operator is to localise the state of that particle about the random position ${\bf z}_t$. The particles are assumed to be distinguishable and each particle experiences independent collapses at independent Poisson-distributed random times with rate $\lambda$. The parameter $\alpha$ sets a length scale for the localisation. 

So why is it that we only see collapse during measurements if the collapses are perpetually affecting all the particles? The reason is that during a quantum measurement, a large number of particles in the measuring device (assumed to be something like a rigid pointer) become entangled with the state of the system of interest and it only requires that one constituent particle undergoes a collapse event to cause the entire collapse process. This means that the rate of collapse per particle $\lambda$ can be made sufficiently small that collapses rarely affect small systems but given the large number of particles in the device, the chance of a constituent particle undergoing a collapse in a small time can be sufficiently high. 

For the CSL model we have \citep{ghirardi1990markov} (Sec. IVC)
\begin{align}
\hat{J}[z_t] = W^{1/2}\exp\left\{-\frac{\beta}{2} \int d {\bf x} \left[ \hat{N}({\bf x}) - z_t({\bf x})\right]^2\right\},
\label{CSLJ}
\end{align}
where $W$ is a normalisation factor, $\hat{N}({\bf x})$ is the (smeared) particle number density operator about point ${\bf x}$, and $\beta$ is a collapse strength parameter. These collapses occur with Poisson distributed random times with rate $\mu$. Here $z_t({\bf x})$ is a real random function. The effect of this collapse operator is to sharpen the state of number density at each point in space. This is not a precise projection of the $N({\bf x})$-state; there is some spread determined by the parameter $\beta$ and the timescale of interest. For systems of small number density this spread can be large enough to leave spatial superposition states untroubled; but for systems with large number density (such as measuring devices) a Schr\"odinger cat state of, for example, different macro pointer positions with a large quantum variance in number density would be rapidly suppressed. This collapse operator also has a localisation effect which can be easily understood in the case of a single particle since the only state of well-defined number density (i.e. zero quantum variance) is one that is perfectly localised. Strictly, the CSL model as it is usually presented is the continuous limit of this model which involves taking the rate $\mu$ to infinity whilst letting $\beta\rightarrow 0$ in such a way that $\mu\beta = 2\gamma$ is a constant, and where $\gamma$ is the usual CSL parameter \citep{ghirardi1990markov} (Sec. IVC). For the sake of austerity we shall continue to call this the CSL model. 

At this stage we might be tempted to argue that the wave function represents the state of the world. This seems reasonable in light of the fact that having incorporated collapses into the dynamics, the wave function has been purged of unwanted macro superposition states. With collapse, the dynamical process accurately updates the wave function to reflect previous measurement events.

Ultimately we require that our theory explain the world of our experience which involves spatial arrangements of matter which change with time, or more generally local properties in fixed regions of spacetime \citep{bell1976theory}. But the wave function is not an object that belongs in the 3+1 dimensional spacetime of our experience. It belongs to a large (or perhaps infinite) dimensional Hilbert space. If the wave function represents the state of the world there is some work to do in converting it into definite local properties of the world at fixed points in space and time. One approach is to determine the precise density of matter in space as the quantum expectation of a suitable matter density operator (this is often called the matter density ontology). A problem with this kind of scheme is that it is possible to come up with other characteristics that can be defined precisely in expectation but which would not be directly observable (a simple example being the expected momentum of a well localised particle). For a precise correspondence between the wave function and local properties in spacetime it would seem that we must choose some preferred operator(s) to act as directly experienced features. This might not seem too bad given that we have already chosen a preferred basis for collapse, but the idea that, even though the wave function describes the world, we only have direct access to a seemingly limited feature of it, seems arbitrary.

In practice in the laboratory, the wave function only reflects the state of a system insofar as it accurately predicts measurement statistics. For an experimenter it is the measurement results that are the experienced occurrences. Noting that the description of a generic collapse model presented in equations (\ref{eq:CM1})-(\ref{eq:CM4}) above is formally equivalent in its effect on the wave function to a generalised quantum measurement (see \citet{nielsen2010quantum}), and that in this correspondence the collapse outcomes $z$ play the role of measurement outcomes, leads to the suggestion that the collapse outcomes (rather than operator expectations) should be treated as local properties in spacetime \citep{bell2001there}. This is a far more natural option given that they are a component part of the model, integral to the dynamical process, and that they belong in spacetime. (See \citet{esfeld2014grw,esfeld2014primitive,egg2015primitive} for further discussion.)

In the GRW model the collapse outcomes are points in spacetime representing a sample of individual particle locations. A physical object then corresponds to a collection of collapses in a localised spacetime region. A massive object (i.e.~a dense collection of collapses in a local region) will persist in time since the particles (of the wave function description) will continue to undergo collapses.

In the CSL model the collapse outcomes are assignments to points throughout space and time of values $z$ representing the local number density of particles. This is a much more straightforward ontology---the collapse outcomes are a dense sample of local matter density throughout spacetime. However, it has been pointed out by \citet{bacciagaluppi2010collapse} that in the continuous version, the variance in $z$ is effectively infinite so that it is difficult to see how a smooth distribution of matter is recovered. This is also the case for the discrete-in-time version (\ref{CSLJ}) but the problem can easily be rectified by having the collapses occur at discrete points in space as well as time (see Fig.\ref{F1}). With a view to forming a relativistic model of collapse we can suppose that the CSL collapses are randomly located with uniform distribution in {\it spacetime} volume (up to Poisson-type fluctuations expected to result from the random sprinkling) with density $\mu$, and that each collapse takes the form
\begin{align}
\hat{J}(z_x) = \left(\frac{\beta}{\pi}\right)^{1/4}\exp\left\{-\frac{\beta}{2} \left[ \hat{N}({\bf x}) - z_x\right]^2\right\},
\label{csleqdisc}
\end{align}
where $x = ({\bf x},t)$. This collapse occurs at time $t$ and leads to a narrowing of the (smeared) number density state at position ${\bf x}$. The variance in each $z_x$ is finite and the usual CSL model is recovered by letting $\mu$ tend to infinity whilst $\beta\rightarrow 0$ with $\mu\beta = 2\gamma$ (recall $\gamma$ is the CSL parameter). 

In both these examples, when we gather together many of the realised $z$ values, we find that they provide a representation of where matter is in spacetime. It thus makes sense to regard these $z$s as actually representing the definite matter distribution in spacetime. Since the collapse outcomes correspond to the outcomes of generalised measurements, then empirical predictions of the distribution of matter in the world in terms of $z$ will agree with those of the standard quantum formalism. Small differences are to be expected since the collapse process leads to loss of coherence and heating offering a way to experimentally test for collapse (see \citet{bassi2013models} and references therein).

However, this construction sees the matter distribution as a kind of by-product of the collapse dynamics of the wave function. It is natural to ask whether it is possible to begin with a description in terms of physical events $\{z\}$ in which the role of the wave function is simply to determine the probability distribution of $z$s in the future given information about $z$s in the past. In the next subsection we outline how this can be achieved. In light of this it seems wrong to call $\{z\}$ the collapse outcomes; we should simply regard them as the distribution of matter.

\subsection{Collapse as inference}
\label{S2a}

It may seem that the distribution of matter (defined in terms of the collapse outcomes) has more of an intertwined relationship with the wave function than suggested by the claim that the wave function is simply a state which makes probabilistic interpretations about future variables given known past ones. After all, the collapse outcomes are generated by the physical collapse process of the wave function. But in fact, as we will show here, one can simply regard the wave function as defining a probability distribution and think of a collapse event described by Eq.~(\ref{eq:CM2}), not as a physical process, but as representing an inference. 

In order to make this point consider the following (see also related work in \citet{bedingham2011hidden, tumulka2011comment}). We assume that there is some c-number valued time-dependent quantity $A(t)$ which is not directly observable. In fact the only access to this quantity is via discrete noisy information which takes the form
\begin{align}
z_t = A(t) + B_t,
\end{align}
where $B_t$ is a noise variable assumed to have a Gaussian distribution with standard deviation $\sigma$ so that the distribution of $z_t$ is centred about $A(t)$ with width of order $\sigma$. The variable $A(t)$ can be thought of as an attempt to capture the behaviour of the mean value of $z_t$, the value once it has been stripped of random fluctuations. We further assume that $A$ is `quantum' in the sense that the probability distribution for $A(t)=A$ at time $t$ is given by 
\begin{align}
P_t\left( A\right) = \frac{|\langle A | \psi_t \rangle|^2}{\langle \psi_{t}| \psi_{t} \rangle},
\end{align}
where $|\psi_t\rangle$ is a Schr\"odinger-picture wave function satisfying the Schr\"odinger equation, and where possible values of $A$ are represented by an operator $\hat{A}$ with eigenstates $|A\rangle$ satisfying $\hat{A}|A\rangle = A|A\rangle$. Now let us ask how we should update this probability distribution on the basis of the noisy information $z_t$ at time $t$. Using Bayes' theorem we have
\begin{align}
P_t\left( A| z_t\right)= \frac{P_t\left( A\right)P_t\left( z_t|A\right) }{P_{t}\left( z_t\right)}.
\end{align} 
The probability distribution for $z_t$ given $A$ is a Gaussian of width $\sigma$ centred on $A$ so that
\begin{align}
P_{t}\left( z_t|A\right) = \frac{1}{(2\pi\sigma^2)^{1/2}}e^{-(A-z_t)^2/2\sigma^2},
\end{align}
and the probability distribution for $z_t=A+B_t$ is a convolution of the probability distributions of $A$ and $B_t$,
\begin{align}
P_{t}\left( z_t\right) = \int dA \frac{1}{(2\pi\sigma^2)^{1/2}}e^{-(A-z_t)^2/2\sigma^2} \frac{|\langle A | \psi_t \rangle|^2} {\langle \psi_{t}| \psi_{t} \rangle}.
\label{eq:pz}
\end{align} 
This means that we can write
\begin{align}
P_{t}\left(A|z_t\right) = \frac{|\langle A | \psi_{t+} \rangle|^2}{\langle \psi_{t+}| \psi_{t+} \rangle},
\end{align} 
where 
\begin{align}
|\psi_{t+}\rangle = \frac{1}{(2\pi\sigma^2)^{1/4}}e^{-(\hat{A}-z_t)^2/4\sigma^2} |\psi_{t}\rangle.
\end{align}
This is the standard dynamical collapse process for a jump operator of the form
\begin{align}
\hat{J}(z_t) = \frac{1}{(2\pi\sigma^2)^{1/4}}e^{-(\hat{A}-z_t)^2/4\sigma^2},
\end{align}
and this is the form taken in the GRW model (with an $\hat{A}$ operator for each component of the position operator for each particle). The collapse process is mathematically equivalent to a Bayesian updating of the probability distribution on the basis of noisy information about the precise value of the variable $A$. Furthermore, the unconditioned probability for the variable $z_t$ given in Eq.(\ref{eq:pz}) is precisely of the form of the standard collapse model probability 
$P_{t}\left(z_t\right) = \langle \psi_{t+}|\psi_{t+}\rangle/\langle \psi_{t}|\psi_{t}\rangle$.

The CSL model of Eq.(\ref{CSLJ}) can be treated in a similar way and we arrive at the same conclusion by choosing a noisy information field of the form
\begin{align}
z_t({\bf x}) = A({\bf x}) +B_t({\bf x}),
\end{align}
where the field variable $A({\bf x})$ represents the (unsmeared) number density of particles, and $B_t({\bf x})$ is a field of Gaussian noise with spatial correlations described by a covariance function which is related to the smearing function of the CSL model. The noise field expressed here $z_t(\bf x)$ is a linear transformation of that of the CSL model of Eq.(\ref{CSLJ}). 

\begin{figure}[h]
        \begin{center}
        	\includegraphics[width=\textwidth]{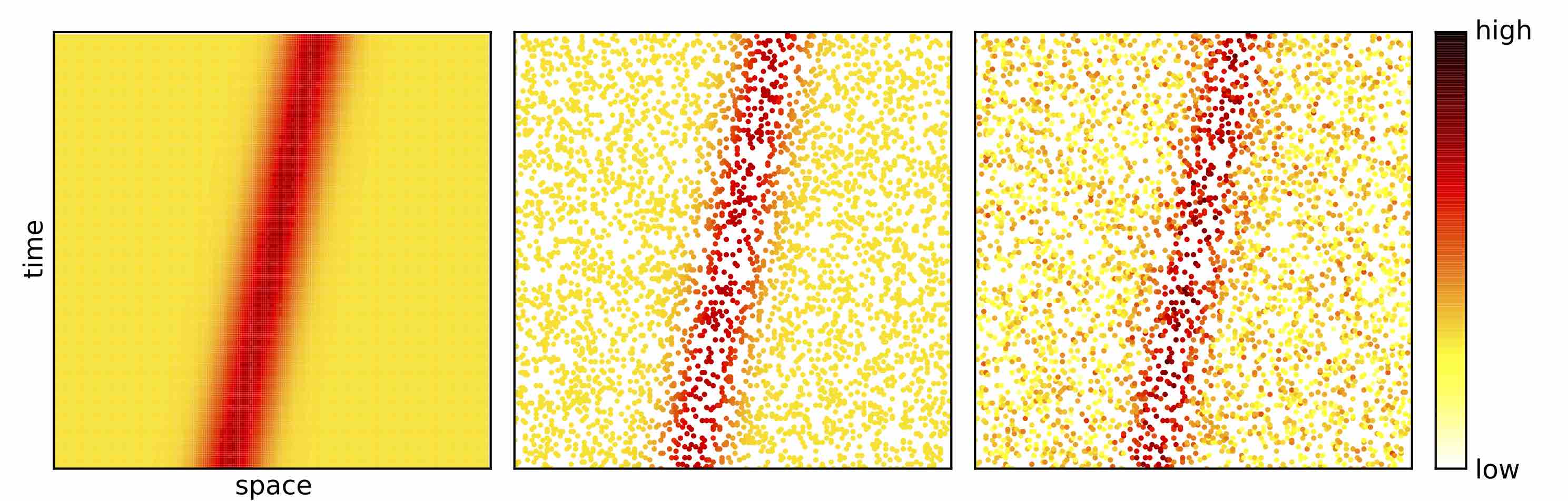}
        \end{center}
\caption{ A comparison of pictures of matter density in spacetime. The left panel displays an object with Gaussian distribution of matter density in space moving with constant velocity to the right. The central panel shows a pointillistic view of the same matter density for a Poisson-distributed sample of points in spacetime. The right panel shows the same points but where the matter density values now include a Gaussian noise component---this picture is the natural ontology of the CSL model as expressed in Eq.\ref{csleqdisc}.}
\label{F1}
\end{figure}

The CSL model of Eq.(\ref{csleqdisc}) is treated in the same way as the GRW model above where now $A$ represents the smeared number density at each randomly sprinkled point $x = ({\bf x}, t)$ (see Fig.\ref{F1}).

Note that a choice of initial wave function at the beginning of the Universe is essentially a choice of the initial probability distribution for $A$. Supposing that this is an objective chance (perhaps obtained via a best-system analysis) makes the wave function determinate (up to an overall phase) at every time slice provided that we always update for past $z$s. The unitary dynamics of the wave function defines the probability distribution for $A$ at future times and it is here that quantum behaviour enters. 

In practice we do not know the initial state of the Universe and we are unable to keep up to date with every single past value of $z$. However, if we prepare a wave function for a small system on the basis of a suitable measurement we can provide a good approximation to the wave function for performing quantum mechanics in practice. A wave function for which we have ignored the results of previous collapse outcomes (most likely we can only perform the update approximately when we are able to amplify the effects in a quantum measurement) is still a perfectly valid tool for estimating the value of $A$, it simply rests on incomplete information. Furthermore, there are two clear examples of when a collapse can be ignored: (i) when it happens in a distant region where the quantum state is uncorrelated with the system of interest; and (ii) for small systems where the width of the noise distribution $\sigma$ (equivalently the width of the collapse operator) is much greater than the uncertainty in $A$ determined by the wave function.

\subsection{Time reversal symmetry}
\label{S2b}

The picture of dynamical collapse presented above appears to be inherently time asymmetric. More generally the idea of objective collapse of the wave function is at odds with time symmetry since the wave function at any point in time is shaped by those collapse events that have happened in the past but not in the future (see also \citet{penrose2004road} Ch.~30.3). Here we will show that there is an underlying time-reversal-symmetric structure \citep{bedingham2015time,bedingham2016time}.

Consider the entire history of the Universe from the initial state $\hat{\rho}_I$ (which we generalise to a density matrix state) at the beginning of time in the past, all the way to a final condition $\hat{\rho}_F$ at the end of time in the future (we treat the final condition as a POVM on the final state; there is no demand that the final state be precisely of the from $\hat{\rho}_F$ and we can have $\hat{\rho}_F\propto \hat{\mathbb{1}}$ expressing no constraint). By repeatedly using the probability rule (\ref{eq:CM3}) we can determine the probability of a given set of time-ordered collapse outcomes $\{z_1,z_2,\ldots,z_{n}\}$ (of which we assume that there are $n$ of them) occurring at a given set of time intervals $\{\Delta t_0,\Delta t_1,\ldots,\Delta t_{n}\}$ ($\Delta t_0$ is the time interval between the start of the Universe and the first collapse event; $\Delta t_1$ is the time interval between the 1st and 2nd collapse events, etc; $\Delta t_n$ is the time interval between the last collapse event and the end of the Universe). We further condition this probability on the initial and final states of the Universe. Given some simple constraints (see below) it turns out that this probability is exactly the same as the probability of the reverse sequence of collapse outcomes $\{z_{n},\ldots,z_2,z_1\}$ with the reversed set of time intervals $\{\Delta t_{n},\ldots,\Delta t_1,\Delta t_0\}$ given the initial condition $\hat{\rho}_F^*$ and final condition $\hat{\rho}_I^*$ \citep{bedingham2016time}, or
\begin{align}
P&(\{z_1,z_2,\ldots,z_{n}\}|\{\Delta t_0,\Delta t_1,\ldots,\Delta t_{n}\}; \hat{\rho}_I; \hat{\rho}_F)
\nonumber\\
&= P(\{z_{n},\ldots,z_2,z_1\}|\{\Delta t_{n},\ldots,\Delta t_1,\Delta t_0\}; \hat{\rho}_F^*; \hat{\rho}_I^*).
\end{align}
A sufficient condition for this to be the case is that there exists a basis $\{|\phi_i\rangle\}$ in which 
\begin{align}
\langle \phi_i | \hat{U}(t) |\phi_j\rangle^* &= \langle \phi_i | \hat{U}(-t) |\phi_j\rangle;\nonumber\\
\langle \phi_i | \hat{J}(z) |\phi_j\rangle^* &= \langle \phi_i | \hat{J}^{\dagger}(z) |\phi_j\rangle.
\end{align}
For a time-independent Hermitian Hamiltonian this means that there exists a basis is which both $\hat{U}$ and $\hat{J}$  are symmetric (this can be shown to be the case for both the GRW and CSL models where the basis is that to which the collapse occurs \citep{bedingham2016time}). The existence of such a basis allows us to define $\hat{\rho}^*$ by
\begin{align}
\langle \phi_i |\hat{\rho}^*|\phi_j\rangle = \langle \phi_i |\hat{\rho}|\phi_j\rangle^*.
\end{align}

So at the level of collapse outcomes $\{z_i\}$ there is time symmetry. The stochastic laws determining the probability of a given complete set of collapse outcomes can be used in either time direction and in both cases they give the same probability. From a dynamical point of view it doesn't matter which direction of time we use. They both give the same result. Assuming that the collapse outcomes alone (without the wave function) give an empirically adequate description of the world then we can conclude that collapse models are time symmetric.

In terms of the wave function things indeed look time asymmetric. But this is simply because the wave function is a way of encoding and conditioning on all the past collapse events. For example, given the initial condition $\hat{\rho}_I$ and the set of all past collapse outcomes $\{z_i|i<j\}$ at time $t_j$, then the quantum state is determined and is given by
\begin{align}
\hat{\rho}_j = &  \hat{U}(\Delta t_j)\hat{J}({z}_{j}) \hat{U}(\Delta t_{j-1}) \cdots \hat{U}(\Delta t_{1})\hat{J}({z}_{1})\hat{U}(\Delta t_0)\hat{\rho_I}
\nonumber \\
& \hat{U}^{\dagger}(\Delta t_0)\hat{J}^{\dagger}({z}_{1})\hat{U}^{\dagger}(\Delta t_1) \cdots \hat{U}^{\dagger}(\Delta t_{j-1})\hat{J}^{\dagger}({z}_{j})\hat{U}^{\dagger}(\Delta t_j).
\end{align}
This encapsulates the maximal information about the past necessary to predict future collapse events. The wave function is time asymmetric for the simple reason that it is defined in a time asymmetric way. 

We can just as easily define a backwards-in-time quantum state from the final condition of the Universe and all collapse outcomes to the future. This would be suitable for estimating past collapse outcomes. We naturally make use of the forward-in-time picture for the reason that we can keep records of the past but not the future. This can be put down to the special initial condition of the Universe (one that is essentially a clean slate upon which to chalk). Indeed it is the imposition of time asymmetric initial and final conditions of the Universe that must be responsible for observed asymmetries predicted by collapse models---these include average energy increase.

The argument presented in this section supports the conclusion that we should treat the collapse outcomes as the primitive elements of collapse models. At the level of collapse outcomes the dynamics are fundamentally time symmetric. At the wave function level it is natural that the description is time asymmetric since the wave function is an asymmetrically derived concept representing  a state of information about past collapse events.

\section{Relativistic structure}
\label{S3}

\begin{figure}[h]
        \begin{center}
        	\includegraphics[width=0.66\textwidth]{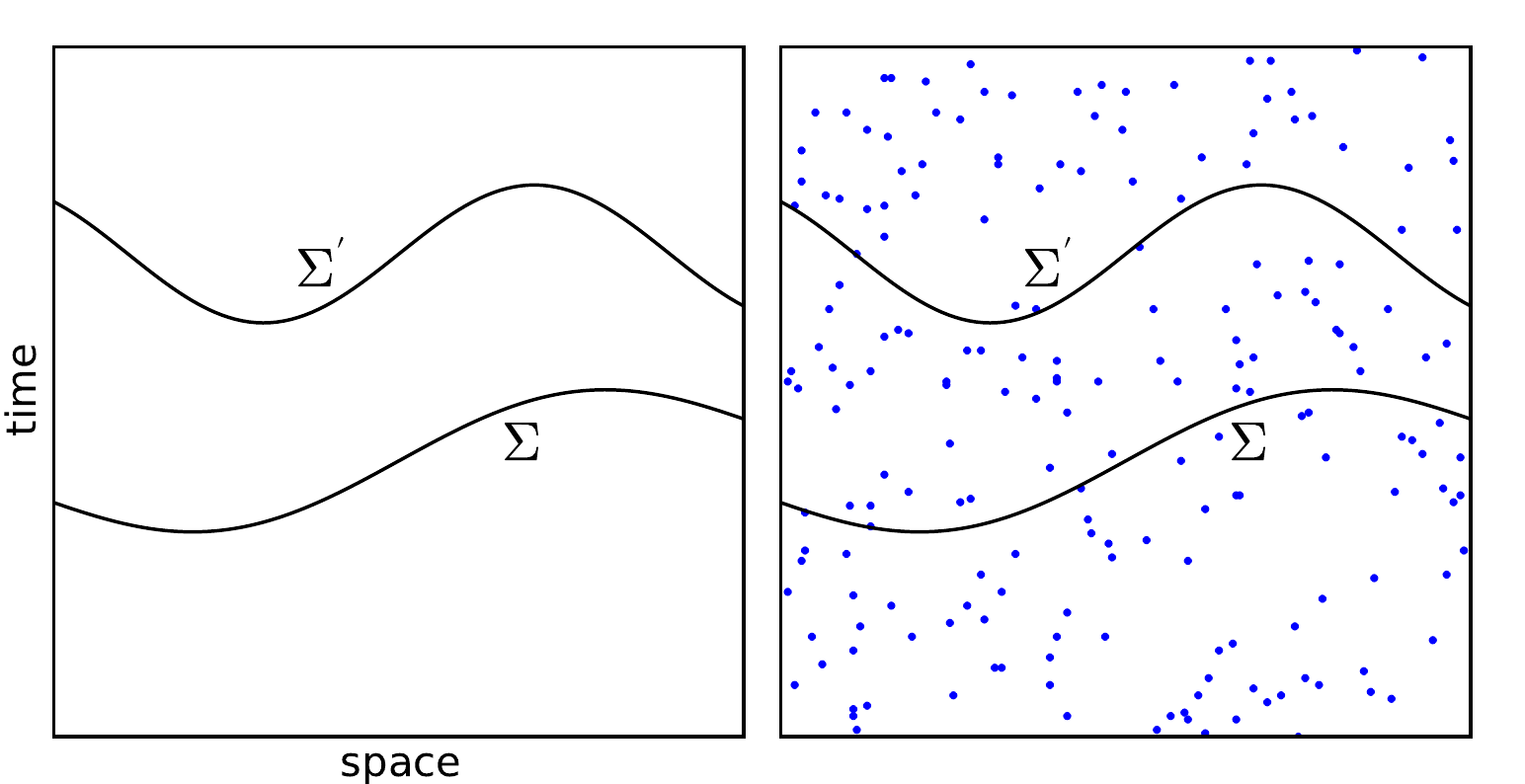}
        \end{center}
\caption{Left panel: spacetime diagram indicating spacelike hypersurfaces $\Sigma$ and $\Sigma'$ with $\Sigma\prec\Sigma'$. Right panel: includes collapse events randomly sprinkled in proportion to spacetime volume using a Poisson process. }
\label{F2}
\end{figure}

The non relativistic structure presented in the previous section can easily be generalised to one that is explicitly relativistic \citep{bedingham2011relativistic,bedingham2011relativistic2,PhysRevD.94.045009}. Earlier, our use of a time parameter $t$ defined a specific time slice on which the state was specified. In general we would like to be able to consistently specify our state with reference to a time slice $t'$ in a different frame, or to specify the state on an arbitrary spacelike hypersurface $\Sigma$. The interaction picture \citep{tomonaga1946relativistically} allows us to do this: to unitarily evolve the state from a surface $\Sigma$ to a new surface $\Sigma'$ (such that $\Sigma \prec \Sigma'$, meaning that no points in $\Sigma'$ are to the past of $\Sigma$, see Fig.\ref{F2}) we use
\begin{align}
|\Psi_{\Sigma'}\rangle  = \hat{U}[\Sigma',\Sigma]|\Psi_{\Sigma}\rangle,
\end{align}
where the unitary operator is
\begin{align}
\hat{U}[\Sigma',\Sigma] = T\exp\left[ -i \int_{\Sigma}^{\Sigma'} \hat{\cal H}_{\rm int}(x) dV\right],
\end{align}
and where $\hat{\cal H}_{\rm int}$ is the interaction Hamiltonian (a scalar operator, ensuring covariance), $V$ is the spacetime volume measure, and $T$ is the time ordering operator. There are many different foliations of spacetime that include both $\Sigma$ and $\Sigma'$ and to ensure that the final result does not depend to the particular choice we use to get from $\Sigma$ to $\Sigma'$, we must impose the microcausality condition: $[\hat{\cal H}_{\rm int}(x),\hat{\cal H}_{\rm int}(y)]=0$ for spacelike separated $x$ and $y$. An easy way to see why this must be true is to note that if we advance the hypersurface first at $x$ and then at $y$ we should get the same result as when we do this in the opposite order. The microcausality condition ensures that this works out.

Now we add the collapse events. These are associated to spacetime points. By randomly sprinkling the locations of the events in proportion to spacetime volume using a Poisson process (see Fig.~\ref{F2}) we ensure that the events are distributed in this way in all frames of reference \citep{bombelli2009discreteness}. The construction is therefore Lorentz invariant. We assume that as the hypersurface $\Sigma$ crosses a point $x$ to which a collapse event is associated the state changes spontaneously according to
\begin{align}
|\Psi_{\Sigma}\rangle \rightarrow |\Psi_{\Sigma^+}\rangle  = \hat{J}_x(z_x) |\Psi_{\Sigma}\rangle,
\end{align}
where $\hat{J}_x$ is a scalar operator (for covariance). The random variable $z_x$ has the probability distribution
\begin{align}
P(z_x|\Psi_{\Sigma}\rangle = \frac{\langle \Psi_{\Sigma+}|\Psi_{\Sigma+}\rangle}{\langle \Psi_{\Sigma}|\Psi_{\Sigma}\rangle}
=\frac{\langle \Psi_{\Sigma}|\hat{J}^{\dagger}_x(z_x)\hat{J}_x(z_x)|\Psi_{\Sigma}\rangle}{\langle \Psi_{\Sigma}|\Psi_{\Sigma}\rangle},
\end{align}
and the collapse operator $\hat{J}_x$ should satisfy the completeness property
\begin{align}
\int dz \hat{J}^{\dagger}_x(z)\hat{J}_x(z) = \hat{\mathbb{1}}.
\end{align}
Then, as with the non-relativistic case in Section \ref{S2}, the construction is analogous to a generalised quantum measurement ($\hat{J}_x$ corresponds to a generalised measurement operator, $z_x$ corresponds to the outcome). Furthermore, the results of sections \ref{S2a} (collapse process understood as Bayesian inference) and \ref{S2b} (time reversal symmetry of collapse outcomes) are straightforwardly generalised to this relativistic scheme.

In the relativistic setting, for the same reason that $\hat{\cal H}_{\rm int}$ must satisfy the microcausality condition, the collapse operators must also satisfy further microcausality conditions
\begin{align}
[\hat{J}_x(z_x),\hat{J}_y(z_y)]= 0
\quad\text{and}\quad
[\hat{J}_x(z_x),\hat{\cal H}_{\rm int}(y)]= 0,
\end{align}
for spacelike separated $x$ and $y$. As demonstrated in detail in \citet{PhysRevD.94.045009}, with these conditions, it follows that: (i) given the set of collapse outcomes $\{z_x\}$ for all collapse events at locations $\{x|\Sigma\prec x\prec \Sigma'\}$ then the final state on $\Sigma'$ is unambiguously defined by the dynamical process given the initial state on $\Sigma$, in particular it is independent of the foliation used to get from $\Sigma$ to $\Sigma'$; and (ii) the probability of a given set of collapse outcomes at given locations $\{x|\Sigma\prec x\prec \Sigma'\}$ is independent of the foliation used to get from $\Sigma$ to $\Sigma'$.

Therefore, the procedure by which we predict the set of collapse outcomes in a spacetime region is covariant and does not depend on a particular frame or foliation. The probabilities for collapse events in a given region as detailed in the previous paragraph may depend on the hypersurfaces $\Sigma$ and $\Sigma'$, however $\Sigma$ is necessary in order to specify the initial state upon which the probabilities are conditioned, and $\Sigma'$ is used simply to specify the region of spacetime which is of interest. By contrast, in order to invoke a state history it is necessary to choose a particular foliation through which the state evolves. The state histories for two different foliations might be very different. In particular if the states of spatially separated regions are entangled, the local state description for one of these regions can depend significantly on the particular leaf on the foliation passing through it. This does not alter the fact that collapse events are consistent and predicable in a way that is independent of foliation.

It is the collapse outcomes which are invariantly specified and it is the collapse outcomes that tie together the different state histories on different foliations such that they are really just different descriptions of the same events \citep{myrvold2002peaceful}. Furthermore treating the quantum state as a state of information about past collapses makes it perfectly natural that it should be dependent on a choice of hypersurface since this is needed to demarcate the past that we are talking about. Relativistic considerations therefore further boost our conjecture that it is the collapse outcomes that are to be seen as primitive and the quantum state as a derivative concept.

It remains only to find a form for $\hat{J}_x$ which satisfies these constraints (scalar + microcausality). For a complex scalar field for example, a well motivated choice is \citep{ghirardi1990relativistic}
\begin{align}
\hat{J}_x(z_x) = \left(\frac{\beta}{\pi}\right)^{1/4}\exp\left\{-\frac{\beta}{2}(\hat{\phi}^{\dagger}(x)\hat{\phi}(x) - z_x)^2\right\},
\label{eq:relJ}
\end{align}
[cf.~Eq.~(\ref{csleqdisc})]. The effect of this collapse operator is to focus the quantum amplitude associated to the modulus squared of the field operator about the randomly chosen value $z_x$. Without the unitary dynamics we would expect many such collapse events to gradually bring the state to an eigenstate of $\hat{\phi}^{\dagger}(x)\hat{\phi}(x)$. In fact the effect of the free unitary dynamics is to disperse the state in competition with the collapse so that typically we would find a stable state with a non zero variance in the field value.

It is well known that the choice of collapse operator (\ref{eq:relJ}) is problematic \citep{ghirardi1990relativistic}. It is to be expected that collapse models lead to an increase in average energy: for realistic models the collapses lead to localisation and by the uncertainty principle this results in momentum dispersion and therefore average energy increase. The problem with the present model is that the rate of increase in the energy density is infinite. If $\hat{H}$ is the standard free complex scalar field Hamiltonian then the expected change in energy for a collapse event occurring at $x$ is
\begin{align}
\Delta E = \int dz\frac{\langle\Psi_{\Sigma}|\hat{J}_x(z)[\hat{H},\hat{J}_x(z)]|\Psi_{\Sigma}\rangle}{\langle\Psi_{\Sigma}|\Psi_{\Sigma}\rangle} = \delta^{3}(0) \frac{\beta}{2}\langle\hat{\phi}^{\dagger}(x)\hat{\phi}(x)\rangle,
\end{align}
(in the limit in which $\beta$ is small). A simple resolution follows from the assumption that spacetime is fundamentally discrete. Even for classical spacetimes there are approaches to doing this in a Lorentz invariant manner \citep{bombelli1987space}. The point is that the discreteness sets a fundamental length scale, $a$, which means that $\delta^{3}(0) \sim a^{-3}$, rendering the energy increase finite. In remains to set the parameters of the model in such a way that energy increase is small yet collapse effects are significant enough to provide an explanation of the macro world of our experience. Furthermore we are not necessarily constrained by the fact that $a$ might be the Planck scale. If for the collapse model to work a larger length scale is necessary then within a discrete spacetime it is conceptually unproblematic to consider a quantum field which only exists on a coarse subset of the spacetime lattice. If only this field experiences collapses then other fields would collapse if they had direct interaction with this field. (We also note that there is an alternative resolution to the problem of infinite energy increase making use of an unusual ``off-mass-shell" quantum field to mediate the collapse process \citep{bedingham2011relativistic,bedingham2011relativistic2,PhysRevD.94.045009}.)

For multiple quantum fields we can replace $\hat{\phi}^{\dagger}(x)\hat{\phi}(x)$ in the collapse operator (\ref{eq:relJ}) by a sum of equivalent ``mass" terms for each of the fields. Another option could be that each individual field undergoes its own individual collapse process, or else that only one field undergoes collapse and others collapse as a result of their unitary interaction and subsequent entanglement with this field.

In summary, the lessons for collapse models from relativity are a consolidation of the idea that it is the collapse outcomes that form the primitive ontology and further that spacetime has a discrete structure. In the next section we will try to strengthen the case for these features by arguing that the problems of incorporating gravity might also be resolved by taking this view.

\section{Including gravity}
\label{S4}

A remarkable feature of dynamical collapse, and in particular the picture of collapse presented here with primitive status given to the collapse outcomes, is that it undermines the main arguments for the need for a quantum theory of gravity and that it may enable the formation of a self-consistent theory of classical gravity and quantum matter (see also \citet{tilloy2016sourcing}). The starting point for any attempt to do this is to try to make sense of Einstein's equation
\begin{align}
G_{\mu\nu} = 8\pi T_{\mu\nu},
\end{align}
where, given that the left side must be a c-number valued tensor, the question is what to use for the right side. We cannot simply use the quantum operator for the energy momentum tensor and so the usual approach is to use the expectation of the quantum matter energy momentum operator. For collapse models the more natural choice is to build the energy momentum tensor from the collapse outcomes which represent the distribution and flow of matter through spacetime. This will clearly be a challenge, and we offer no clear route to doing this here. However, we will analyse some objections to the semi-classical approach to treating gravity and point out how they are avoided by this picture.

The first two issues are due to \citet{eppley1977necessity}. The first arises when it is assumed that a gravitational wave can interact with quantum matter without causing the state to collapse. In a simplified version  which captures the essential idea, a particle is prepared in a superposition of two spatially separated localised states, each in its own box \citep{carlip2008quantum}. Then, by monitoring one of the boxes by scattering gravitational waves from the particle we can, if the gravitational wave does not collapse the particle state, observe the particle as belonging to a superposition state and detect a change if a measurement is performed on the other box to determine if the particle is located there. This enables signals to be sent at speeds greater than the speed of light.

The reason why this argument does not work for gravity sourced by the collapse outcomes is due to the fact that in order for there to be any influence from matter on gravity there must be collapses. Only those collapses afflicting the box being monitored will influence the gravitational waves being used to monitor it. The no-signalling theorem can then be used to show that any measurement on one box cannot influence the collapse outcome statistics for the other box and therefore cannot influence the scattering amplitude of a gravitational wave.

The second issue of \citet{eppley1977necessity} concerns cases when interaction between a gravitational wave and some matter requires the collapse of the matter state. A gravitational wave could in principle be used to measure the location of some matter prepared in a well-defined momentum state. Since the gravitational wave could have arbitrarily small momentum so that the momentum state of the matter is undisturbed, then both the position and momentum could be known with accuracy. 

This argument works on the premise the momentum is conserved. This is simply not true for realistic collapse models, and if the state becomes localised by any means (including interaction with a gravitational wave) then by its wavelike nature, its momentum becomes uncertain.

Further arguments are made by \citet{page1981indirect} who point out that if there is no collapse of the wave function, and if gravity is sourced by the expectation of quantum matter energy momentum, then the theory actually contradicts experimental results since it predicts gravitational effects sourced by quantum matter on branches of the wave function other than ours. Their experiment involves quantum mechanical decision and amplification to test gravitational effects from other branches: depending on the decision outcome a large mass is moved and its gravitational field observed. 

This issue is clearly not relevant for collapse models where the energy momentum of the quantum state is continually representative of the distribution of energy and matter that we experience, but, \citet{page1981indirect} point out, any collapse model of the wave function leads to a contradiction of the Einstein equation since the left side is automatically conserved, $\nabla^\mu G_{\mu\nu} = 0$ (the contracted Bianchi identity), yet the right side will not be conserved under a collapse process. It is a generic feature of realistic collapse models that they do not conserve energy momentum. Collapse models are by nature stochastic and as pointed out earlier, the spontaneous localising feature is accompanied by jumps in momentum and energy. Furthermore if we are to consider using the noisy collapse outcomes as the basis for the energy momentum source the problem is compounded. The Einstein equation simply won't work with a stochastic energy momentum source. 

We have already seen in the Section \ref{S3} that discrete spacetime structure offers a possible resolution to the problem of infinite energy increase in relativistic versions of collapse models, there is also an indication that it could resolve this inconsistency of the Einstein equation.

An example of a spacetime with discrete geometry is Regge calculus \citep{regge1961general} in which a manifold is built from elementary simplexes (e.g a 2-dimensional simplex is a triangle; a 3-dimensional simplex is a tetrahedron, etc). Within each simplex the continuous space is flat and we can define coordinates for any two adjacent cells which are flat throughout the two. The result is a Riemannian structure on all but the boundary of the joints of several simplexes. This is where the curvature is concentrated. 

We will only note that the Regge analogue of the contracted Bianchi identity is not precise. In fact only in the limit that the space becomes smooth is it the case that the analogue of the Einstein tensor is conserved \citep{regge1961general}. This opens the door to a consistent equation for discrete spacetime geometry sourced by stochastic and non-conserved matter.

A promising approach to describing discrete spacetime with Lorentz invariance is as a causal set \citep{bombelli1987space}, which is a set of points with a causal order relation $\prec$ such that if $x\prec y$ then the point $y$ is to the causal future of $x$. Discreteness is expressed by the fact that the set is locally finite meaning that between any two causally ordered points $x\prec y$ there is a finite number of points $z$ such that $x\prec z\prec x$. Comparison with a Lorentzian manifold is made using the notion of faithful embedding: a causal set can be faithfully embedded in a Lorentzian manifold if the points of the set can be mapped into the manifold in such a way that the order relation matches the causal ordering of the manifold, and the number of set elements mapped into a region is on average proportional to the volume of the region. It is a central conjecture that a given set cannot be faithfully embedded into two manifolds that are not similar on scales greater than the discretisation scale.

The fundamental structure of causal sets does not demand that they can be embedded in any manifold, and to the extent that a causal set is well approximated by a manifold, it need not satisfy the Bianchi identities precisely. Given a causal set that can be faithfully embedded in a Lorentzian manifold one can always add and take away points and order relations [provided that new order relations do not violate the fundamental axioms: antisymmetry ($x\prec y\prec x\implies x=y$), and transitivity ($x\prec y \prec z\implies x\prec z$)], allowing for stochastic deviations of the causal structure and local volume measure with respect to the manifold. The problem for causal sets is rather how the correspondence with a 4-dimensional manifold should arise from all possibilities.

\section{Summary}
\label{S5}

We have argued for an understanding of collapse models in which the substance of the world is definite-valued matter density distributed in space and time and where the wave function describes the patterns which relate matter densities at different points. We have shown that the wave function can be understood as a probability distribution for the future matter density conditional on matter density in the past and in which the collapse process corresponds to a Bayesian update on the basis of new matter density information.

We have seen that the distribution of matter density can be explained in a way that is structurally time symmetric and that the time asymmetry of the wave function dynamics is naturally understood as resulting from an asymmetry in the way that the wave function is defined as a conditional probability distribution. Similarly the probability of a given distribution of matter density can be specified with Lorentz invariance and the foliation dependence of a given state history is a natural consequence of the need to demarcate a past-future boundary for the purpose of conditioning on past data.

Finally we have argued that this picture of quantum matter undermines some key arguments in favour of the need for a quantum theory of gravity, raising the possibility of a consistent theory of quantum matter and classical gravity.

\section*{Acknowledgements}
I would like to thank Chris Timpson, Owen Maroney, and Harvey Brown for useful discussions and comments on an earlier draft.

\bibliographystyle{agsm}
\bibliography{DJBbib}

\end{document}